\documentclass[journal,12pt,onecolumn,draftclsnofoot,]{IEEEtran}

\usepackage{algorithm}
\usepackage{algpseudocode}

\usepackage[utf8]{inputenc}
\usepackage{dsfont}
\usepackage{multirow}
\usepackage[mathscr]{euscript}
\usepackage{amsmath}
\usepackage{amssymb}
\usepackage{amsfonts}
\usepackage{amssymb}
\usepackage{amsmath}
\usepackage{verbatim}
\usepackage[T1]{fontenc}
\usepackage[utf8]{inputenc}
\usepackage{latexsym}
\usepackage{enumerate}
\usepackage{indentfirst}
\usepackage{calligra}
\usepackage{ulem}
\usepackage{graphicx}
\usepackage{array}
\usepackage{float}
\usepackage[colorlinks,citecolor=magenta,urlcolor=blue,linkcolor=blue]{hyperref}
\usepackage{color,soul}
\usepackage{array}

\usepackage{titlesec}
\titleformat{\paragraph}
{\normalfont\normalsize\bfseries}{\theparagraph}{1em}{}
\titlespacing*{\paragraph}
{0pt}{3.25ex plus 1ex minus .2ex}{1.5ex plus .2ex}

\usepackage{authblk}

\usepackage{tikz,lipsum,lmodern}
\usepackage[most]{tcolorbox}

\usepackage{booktabs}
\usepackage{multirow}
\usepackage[colorlinks]{hyperref}
\usepackage[official]{eurosym}

\graphicspath{./assets/}

\usepackage{geometry}
\begin{document}

\title{Impact of Dynamic Operating Envelopes on Distribution Network Hosting Capacity for Electric Vehicles}

\author{Hossein Fani$^{1}$, Md. Umar Hashmi$^{1}$, Emilio J. Palacios-Garcia$^{1,2}$, Geert Deconinck$^{1}$
\thanks{Corresponding author email: hossein.fan@kuleuven.be}
\thanks{$^{1}$KU Leuven, division Electa \& EnergyVille, Genk, Belgium}
\thanks{$^{2}$University College Cork, Cork, Ireland}
}


\maketitle

\begin{abstract}
The examination of the maximum number of electric vehicles (EVs) that can be integrated into the distribution network (DN) without causing any operational incidents has become increasingly crucial as EV penetration rises. This issue can be addressed by utilizing dynamic operating envelopes (DOEs), which are generated based on the grid status. While DOEs improve the hosting capacity of the DN for EVs (EV-HC) by restricting the operational parameters of the network, they also alter the amount of energy needed for charging each EV, resulting in a decrease in the quality of service (QoS). This study proposes a network-aware hosting capacity framework for EVs (EV-NAHC) that i) aims to assess the effects of DOEs on active distribution networks, ii) introduces a novel definition for HC and calculates the EV-NAHC based on the aggregated QoS of all customers. A small-scale Belgian feeder is utilized to examine the proposed framework. The results show a substantial increase in the EV-NAHC with low, medium, and high-daily charging energy scenarios.
\end{abstract}



{\textbf{Disclaimer}: This paper is a preprint of a paper submitted to CIRED 2024 Vienna Workshop and is subject to Institution of Engineering and Technology Copyright. If accepted, the copy of record will be available at IET Digital Library}

\pagebreak

\tableofcontents

\pagebreak

\section{Introduction}
Global warming has emerged as a crucial concern that has significantly impacted the lives of people in the \(21^{st}\) century \cite{globwarmingweb}. European countries have been at the forefront of addressing this worldwide decarbonization through the adoption of extensive low-carbon technologies and electrification efforts across several sectors, including power, heating and transportation \cite{greendeal}. 
Hence, replacing fuel-based vehicles with EVs has drawn considerable interest in recent years. As an example, the Flemish government recently agreed to provide a subsidy of \euro{5,000} on the purchase of new EVs valued below \euro{40,000} \cite{belnews}.
Thus, a significant penetration of EVs into the DN is anticipated. While green and clean vehicles are employed to enhance the sustainability of the grid, they also introduce operational challenges, such as the deterioration of power quality within the system \cite{AnamikaPaper}. As a result, the evaluation of EV-HC has garnered significant attention from researchers and system operators recently.

The EV-HC denotes the number/amount of EVs that can be connected to the DN before observing distribution network incidents (DNIs).
DNI includes line congestion, node voltage violations, and non-compliance with EN 50160 \cite{hashmi2023flexibility}. Although most of the methods used to evaluate EV-HC in the existing body of research, such as deterministic, stochastic, and streamlined approaches \cite{KARMAKER2024113916}, predominantly focus on the examination of passive distribution networks and inflexible charging, the increasing use of active network management in DNs has raised a research question: \textbf{How does the active distribution network affect EV-HC?} 
We aim to 
answer this question in this paper.

Active distribution networks react to system disturbances and impact the dynamic behaviour of the overall power system \cite{conte2022equivalent}.
DOEs are effectively incorporated into the active distribution system by offering viable operational regions that align with \textit{grid} objectives. Increasing the adoption of distributed energy resources (DER) and enhancing their HC is considered one of the most recent DOE's developments in the field of distribution system research \cite{10257504}.

In \cite{MahmoodiPaper}, a predetermined range of actual and reactive power is defined as DOE to distribute the existing network capacity among independent users. Reference \cite{MahmoodiPaper2} proposes a DER hosting capacity envelope, which determines the minimum and maximum HC for a DN. To mitigate the security risk during operation, \cite{BinPaper} proposed a robust DOE that explicitly incorporates both the load and DER production. In addition, they took into account fairness in their distribution of resources. The effects of consumer locational disparity are mitigated completely in \cite{UmarPaper} with the implementation of a Volt-Var control, which inherently incorporates DOEs.

To the best of our knowledge, the studies that investigate the HC of an active distribution system, utilize a definition that has been defined for the passive network. The conventional definition of HC needs to be revamped in the presence of DOEs that provide feasible ranges for operating load and/or distributed generation. Thus, this study suggests a novel EV-NAHC framework that measures the DOEs effect on the EV-HC while considering the aggregated QOS of all EV users.

The literature has established the QoS for EV charging by considering various factors, including personal satisfaction, cost, wait time, and energy demand \cite{RamrajPaper}.
Energy-based QoS, which has been broadly utilized in the literature \cite{HarighiPaper, AbdullahPaper, MelikePaper}, is employed in this study as the DOEs adjust the baseline power to meet the operational characteristics of the system. It is defined as the capability to deliver the required energy to an EV. This capability can also be expressed as the charging time and waiting time for the EV, \cite{KHAKSARI2021102872}, which we don't use in this research.

The contributions to the paper are listed below:
\begin{itemize}
    \item Revamping the EV-HC definition in the active distribution network, which utilizes DOEs to avoid DNIs.
    \item Proposing an EV-NAHC framework that considers aggregated QOS as a HC limiting factor.
    \item Numerically, assessing the EV-HC of three different scenarios for daily EV charging energy and the sensitivity of the proposed EV-NAHC approach to the DOE parameters. We also assess the impacts of DOEs on the EV-HC.
\end{itemize}

The rest of the paper is organized as follows: The proposed EV-NAHC framework is further explained in Section \ref{method}. In Section \ref{results}, numerical case studies are presented in detail. The paper is concluded in Section \ref{conclusion}, which also recommends future work.

\pagebreak

\section{Proposed network-aware HC framework} \label{method}

\begin{figure}[!b]
	\center
	\includegraphics[width=0.93\linewidth]{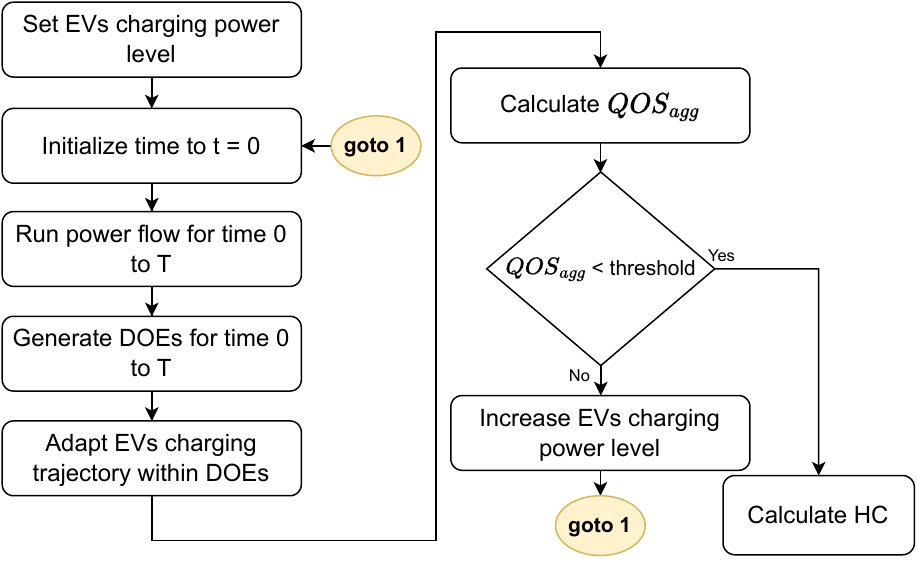}
	\vspace{-8pt}
	\caption{Flow chart for DOE-based EV HC for DNs}
	\label{fig:DOEhc}
\end{figure}

In this section, an EV-NAHC framework, shown in Fig. \ref{fig:DOEhc}, is proposed.
Firstly, the EV profiles utilized in this work are detailed. Secondly, a summary of the DOE generation approach is elaborated. Then, DOE-based EV-HC is summarized. 
The energy-based QoS used to determine the EV-NAHC using DOEs as an active method of network management, is additionally presented in this section.
In this study, the conventional EV-HC, referred to as passive HC, is determined by gradually increasing the charging power of all users until the first network incident occurs.

\subsection{EV charging trajectory data}
To ensure that the EV-HC calculation captures the stochasticity associated with not only the arrival and departure times but also the charging energy, we use real EV charging profiles, which consist of real-world data obtained through the installation of smart meters in a group of 300 households located in Belgium. 
The utilized data extends from Dec-2022 to Dec-2023. 
 
The EV charging sessions, available in the dataset, are detected and characterized using a methodology outlined in \cite{HosseinPaper}. This approach is used to determine the arrival and departure times, daily charging energy, and charging power. In this study, we do not make any assumptions on the flexibility of arrival and departure times, as these times are predetermined based on the historical real data. Subsequently, the EV profiles are categorized into three distinct groups based on their daily charging energy levels: low, medium, and high. The findings will also be provided for the aforementioned three scenarios to examine the potential impact of daily EV charging energy on the EV-HC. 

\subsection{Generation of DOE}
\begin{figure}[!t]
    \centering
    \includegraphics[width=0.75 \linewidth]{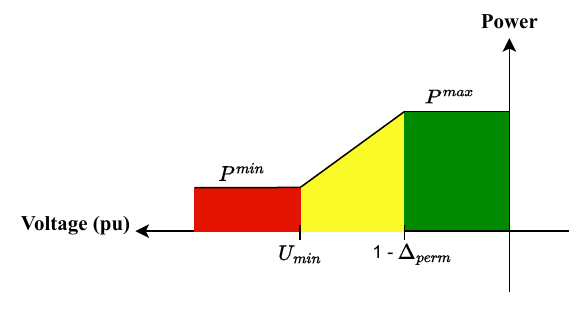}
    \vspace{-14pt}
    \caption{Power control regions of DOE}
    \label{fig:OE}
\end{figure}

In the proposed approach, Fig. \ref{fig:DOEhc}, DOEs are generated based on \cite{10257504}, which utilizes local voltage measurement (or forecast) for estimating the lower and upper levels of active and reactive power within which the flexible devices operated to avoid probable voltage violations. This method assumes that voltage limit violations precede line limit violations, as also observed in \cite{tercca2022deliverable}. Fig. \ref{fig:OE} illustrates the DOE utilized in our study, in which EVs are assumed to solely rely on grid charging without any V2G capability. Thus, in the setting of DOE, only the area comprising undervoltage and positive power is considered. 

In Fig. \ref{fig:OE}, traffic lights are employed in various areas of the DOE according to the voltage thresholds. The green zone represents the voltage-safe area, where the EV is permitted to charge at its maximum power capacity ($P^{max}$). 
The region, denoted by the color red, indicates that the voltage is outside the practical thresholds and entered the restricted zone. Hence, it is essential to decrease the charging power to its minimal threshold, which is delineated as a fraction of the maximum power (Eq. \ref{eq:pmin}). Note that the decentralized DOEs in \cite{10257504}, set $P_{\min} = 0$. However, in the numerical evaluations, it is observed that correction $factor$ should be introduced as the setting $P_{\min} = 0$ leads to over-compensation for voltage regulation.
The yellow area in Fig. \ref{eq:OE} serves as the control zone, where the power is established within the DOE using Eq. \ref{eq:OE}, \ref{eq:OEoptimization}.
$P_{t}^{na}$ is the network-aware EV charging power that results from DOE at each time, which implies adopting the charging trajectory based on the state of the grid. The minimum voltage threshold, denoted as $U_{min}$, is established at 0.9 in accordance with the European Standard EN 50160. $U_{t}$ refers to the local voltage, which is the network state used for generating DOE at each time.
 
\begin{equation}
 P^{min}=factor \times P^{max}
 \label{eq:pmin}
\end{equation}

 \begin{equation}
 P_{t}^{na} \in [P^{min}, \frac{(P^{max}-P^{min})(U_{t}-U_{min}))}{U_{min}-(1-\Delta_{\text{perm}})}+P^{min}]
 \label{eq:OE}
\end{equation}

\begin{equation}
    \begin{aligned}
    \min_{P_{t}^{na}} \quad & \left | P^{max}-P_{t}^{na}\right |\\
    \textrm{s.t.} \quad & P_{t}^{na} \leq \frac{(P^{max}-P^{min})(U_{t}-U_{min}))}{U_{min}-(1-\Delta_{\text{perm}})}+P^{min}\\
      &P_{t}^{na}\geq P^{min}    \\
    \end{aligned}
\label{eq:OEoptimization}
\end{equation}

It is worth noting that the DOE is characterized by two main variables that determine its control region. Various levels of control can be attained by modifying the permissible voltage level (given as $\Delta_{\text{perm}}$) and the factor used for ramping down the charging trajectory (denoted as $factor$, see \eqref{eq:pmin}).

\pagebreak

\subsection{DOE-based EV HC}

The proposed EV-NAHC framework revamps the conventional definition of HC by adjusting the trajectory of EV charging based on DOEs and computing QOS as the limiting factor for HC.
The integration of DOEs in HC calculation would embed grid awareness in EV charging. Fig. \ref{fig:DOEhc} depicts how the framework incorporates DOE-based EV-HC, which utilizes the EV profiles and DOEs previously explained. 

In this framework, after DN and EV level penetration has been set up, power flow is performed to produce voltage profiles. The DOEs are then generated with the aforementioned voltage profiles. By adjusting the baseline state of charging EVs through calculated DOEs, QoS is calculated. Once the QoS reaches the threshold, the HC is estimated; otherwise, the EV penetration level is increased. 

The energy-based QoS for EV charging, which we utilized in our approach, is defined as the divergence in network-aware charging energy consumption from the typical charging trajectory for each customer. Thus, the QoS value is between 0 and 1. This research focuses on the evaluation of network-level HC and identifies aggregate QoS \eqref{eq:aggqosEnergy} for all N EV users as a constraining factor for HC. However, the effects of individual user QoS \eqref{eq:qosEnergy} on the aggregated QoS and the EV-HC are also examined.

\begin{equation}
    QoS_{i} = 1- \frac{ (E^{i} - E_{na}^{i})}{E^{i}}
\label{eq:qosEnergy}
\end{equation}

\begin{equation}
    QoS_{agg} = \frac{\sum_{i=1}^{N}E^{i}-\sum_{i=1}^{N}E_{na}^{i}}{\sum_{i=1}^{N}E^{i}}
    \label{eq:aggqosEnergy}
\end{equation}

\pagebreak

\section{Numerical results} \label{results}

In this section, the proposed approach (EV-NAHC) for assessing the impacts of DOEs on EV-HC is validated on a small-scale feeder in Belgium. To assess the parameters of DOEs multiple scenarios are examined, and the outcomes are presented. Ultimately, the numerical results obtained from implementing the suggested EV-NAHC approach in the specified test feeder are elucidated.

\subsection{Network description (Belgian small-scale feeder)}

\begin{figure}[!b]
  \centering
  \includegraphics[width=0.92 \linewidth]{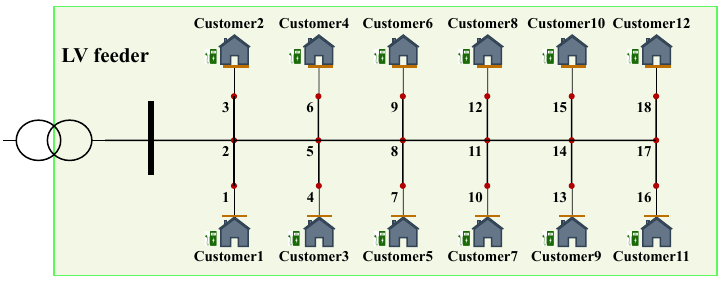}
  \vspace{-6pt}
  \caption{Small-scale Belgian feeder topology}
  \label{fig:19bustopology}
\end{figure}

A typical feeder consisting of 19 nodes from Belgium is employed as a test example. The system consists of a total of 19 nodes and 18 branches, with 12 residential loads being linked to the specified nodes, as depicted in Fig. \ref{fig:19bustopology}. The feeder has been branched from a low-voltage transformer with a capacity of 100 kVA. In terms of topology and parameters, this feeder can serve as a reliable validation tool for assessing the performance of actual Belgian feeders.
The feeder and load profile details of the feeder can be found in \cite{hashmi2021flexible}.

All the loads considered are of residential type. To provide fair treatment to households located at varying distances from the transformer, it is hypothesized that each residence possesses a charging infrastructure. Furthermore, given that the research focuses only on EVs and aims to replicate the most serious effects, it is assumed that none of the households have rooftop solar panels. As a result, the combined value of the EV load and the total load of other appliances represents the load of each household.

To calculate the EV-HC, power flow is performed using OpenDSS. The evaluation process is conducted within a one-day time horizon, with a time interval of 15 minutes. Since the impact of these loads on the EV-HC is not within the scope of this study, it is assumed that the fixed load of every household remains constant throughout various daily energy charging level scenarios.

\pagebreak

\subsection{Case study 1: Fixing DOEs parameters ($\Delta_{\text{perm}}$ and $factor$)}

The region of control of DOEs is limited by altering the values of $\Delta_{\text{perm}}$ and $factor$. Thus, different levels of controllability could be introduced at the consumer's end. In the following, we demonstrate the impact of these parameters on the aggregated QOS and EV-NAHC, and subsequently adjust them according to the results for the next case study. The range of $\Delta_{\text{perm}}$ is defined as [0, 0.1] with a step of 0.01. Three distinct values are assigned to the $factor$, namely 0, 0.2, and 0.5. The aggregated QoS limit is set at 0.8.

\begin{figure*}[!t]
    \centering
    \includegraphics[width=\textwidth]{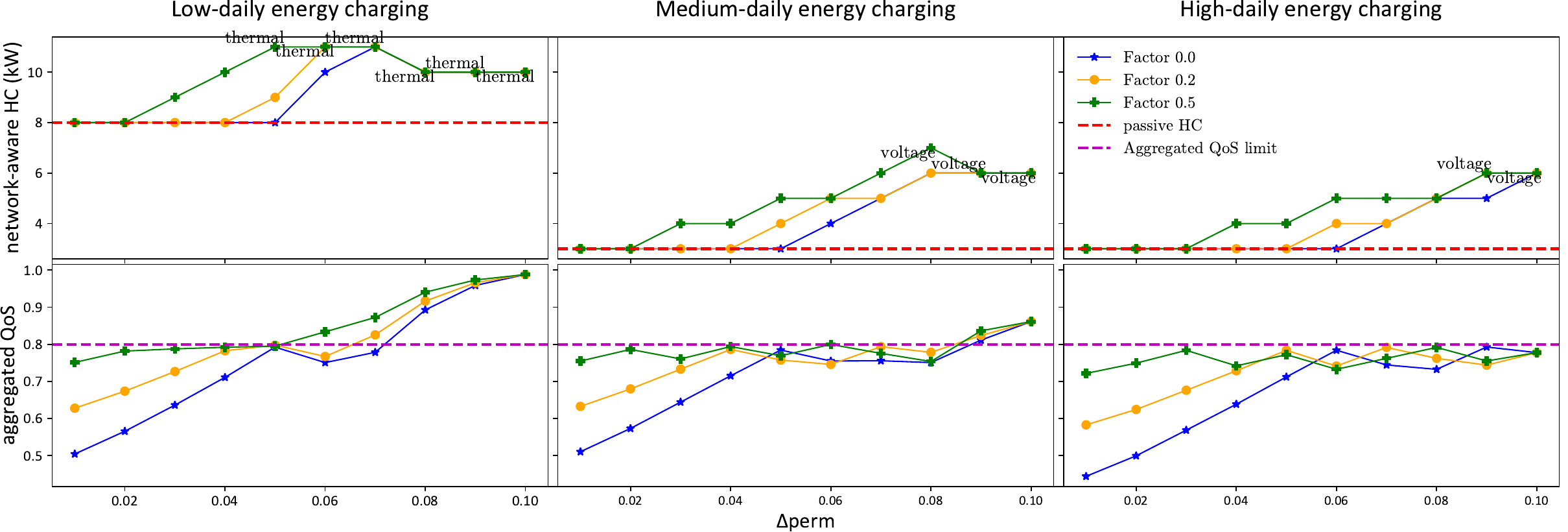}
    \vspace{-8pt}
    \caption{Sensitivity analysis of EV-NAHC and aggregated QoS on ($\Delta_{\text{perm}}$ and factor) for low, medium, and high-daily EV charging energy}
    \label{fig:deltapermanalysis}
\end{figure*}

\vspace{15pt}

\textbf{Sensitivity of aggregated QOS with $\Delta_{\text{perm}}$ \& $factor$:}

The effects of selecting various values of $\Delta_{\text{perm}}$ and $factor$ in creating DOEs on the aggregated QOS and EV-NAHC are depicted in Fig. \ref{fig:deltapermanalysis}. 
The control area of DOE decreases as the values of these parameters increase. Consequently, the aggregated QOS for all consumers improves. However, it could potentially lead to the occurrence of a new network incident.  

Furthermore, Fig. \ref{fig:deltapermanalysis} demonstrates that the effects of $\Delta_{\text{perm}}$ and $factor$ on the aggregated QOS are greater in the low-daily charging energy scenario. This occurs because low-daily energy has shorter charging periods. Hence, their charging trajectory exhibits minimal modifications in comparison to other scenarios. The aggregated QOS for medium and high-daily charging energy scenarios remains relatively stable, as their baseline charging can be considerably adjusted due to the extended duration of their charging sessions.

\vspace{15pt}

\textbf{Sensitivity of EV-NAHC with $\Delta_{\text{perm}}$ \& $factor$:}
In the proposed framework, the EV-HC of a DN would be limited by QoS deterioration and/or unavoided DNIs due to the different generated DOEs.
For instance, when larger values of $\Delta_{\text{perm}}$ are used, the voltage violations that can be controlled are greatly reduced. In that case, DOEs are not able to handle new network incidents that occur, as shown in Fig. \ref{fig:deltapermanalysis}, because EVs are using more power to charge. However, taking into account low values of $\Delta_{\text{perm}}$ can excessively modify the charge profiles and perhaps restrict the QoS for consumers. As a result, the EV-NAHC of DN may not significantly grow due to an unsatisfactory aggregated QoS threshold.

Increasing the $factor$ also shows a similar trend as $\Delta_{\text{perm}}$. 
A high value for $factor$ has already resulted in operational issues for the DN, namely in terms of thermal (for low daily charging energy) and voltage (for medium daily charging) incidents as depicted in Fig. \ref{fig:deltapermanalysis}. Therefore, the EV-NAHC is limited by unavoided DNIs.
This occurs due to the negative correlation between the $factor$ value and the control area of DOEs. 
On the other hand, choosing a smaller value for the $factor$ leads to a significant decrease in the aggregated QOS, and the EV-NAHC is not improved.

In summary, the findings indicate that adjusting the DOE elements is crucial since it simultaneously impacts EV-NAHC and aggregated QOS. The results shown in Fig. \ref{fig:deltapermanalysis} imply that setting $\Delta_{\text{perm}}$ and $factor$ to 0.05 and 0.5, respectively, improves the aggregated QOS and prevents new unavoidable DNIs, which leads to an improvement in EV-NAHC.

\pagebreak

\subsection{Case study 2: Network-aware HC results}

The EV-NAHC results of the suggested framework for the test feeder are presented in Tab. \ref{tab:HCresults}. Based on the outcomes, it can be inferred that the voltage event serves as a limiting factor for the HC in the passive DN across all situations. Hence, employing DOE to address voltage incidents is beneficial.
For the three charging scenarios, i.e. low, medium and high daily energy, the EV-HC improvement due to DOEs is 37.5, 66.7 and 33.3\% respectively. The values shown in Tab. \ref{tab:HCresults} for HC refer to the charging power that a single EV can use to charge the battery. 
As previously stated, the low-daily energy charging EVs in our study lead to a notably larger EV-HC compared to the other two scenarios, mostly due to the reduced charging simultaneity associated with their charging demand.

\begin{table}[!h]
\centering
\scriptsize
\vspace{-6pt}
\caption{\small{EV-HC and EV-NAHC results}}
\renewcommand{\arraystretch}{2}
\resizebox{0.95\columnwidth}{!}{\begin{tabular}{ccccc}
\toprule
\multirow{2}{*}{\textbf{Daily charging energy}} & \multicolumn{2}{c}{\textbf{EV-HC}} & \multicolumn{2}{c}{\textbf{EV-NAHC}} \\ \cmidrule(l){2-5} 
       & HC   & limiting factor   & HC    & limiting factor \\ \midrule
Low    & 8 kW & voltage violation & 11 kW & aggregated QoS             \\
Medium & 3 kW & voltage violation & 5 kW  & aggregated QoS  \\
High   & 3 kW & voltage violation & 4 kW  & aggregated QoS  \\ \bottomrule
\end{tabular}}
\label{tab:HCresults}
\end{table}

\vspace{10pt}

\begin{figure}[!b]
    \centering
    \includegraphics[width=0.75\linewidth]{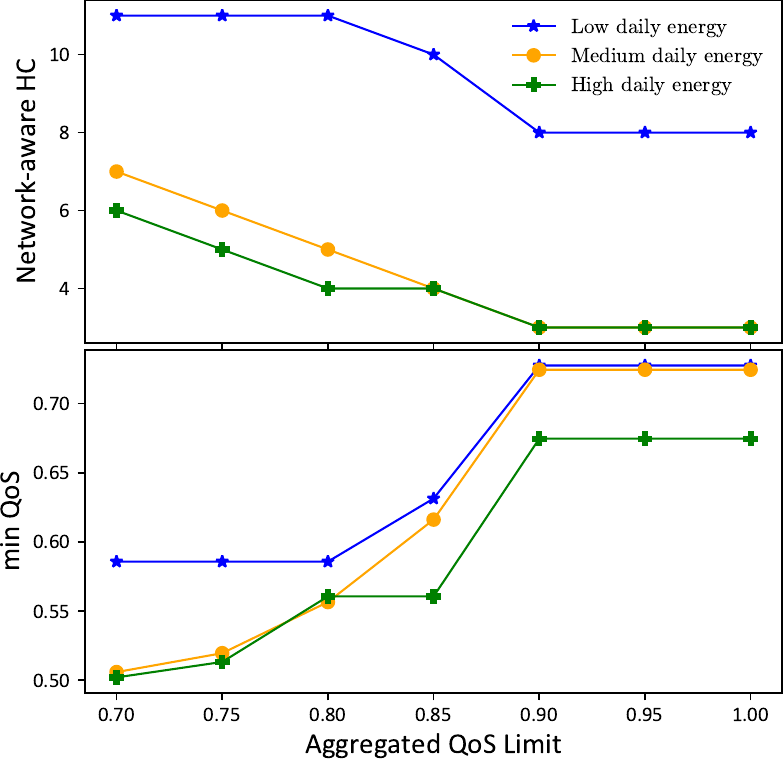}
    \vspace{-8pt}
    \caption{EV-NAHC and minimum QoS Sensitivity on aggregated QoS for low, medium, and high-daily charging energy EVs}
    \label{fig:QoSanalysis}
\end{figure}

\textbf{Sensitivity of EV-NAHC with aggregated QoS limit and minimum QoS:}
Fig. \ref{fig:QoSanalysis} provides a sensitivity of EV-NAHC with different levels of the aggregated QoS threshold. The second plot showcases the implication of aggregated QoS threshold on the worst-case observed QoS level in the network.
As expected, with a reduction in the aggregated QoS threshold, the EV-NAHC is going to increase. The increase for low-daily energy charging profiles is less pronounced compared to medium and high charging profiles.
Due to the radial structure of the DN, the worst-case QoS in the network is substantially lower compared to the aggregated QoS limit.

\vspace{15pt}

\begin{figure}[tp]
    \centering
    \includegraphics[width=0.95 \linewidth]{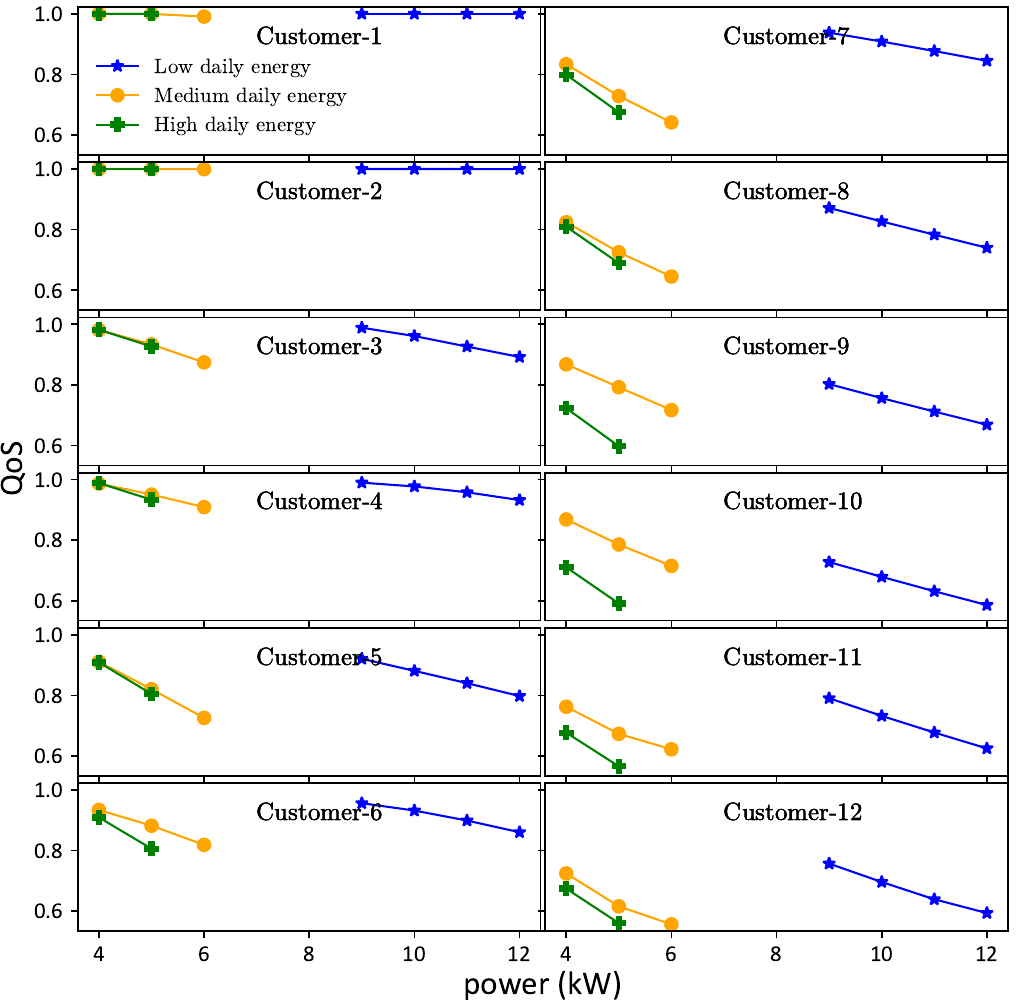}
    \vspace{-8pt}
    \caption{QoS of customers changes in terms of charging power for low, medium, and high-daily charging energy EVs}
    \label{fig:housholdQoS}
\end{figure}

\textbf{Locational variation of QoS:}
Fig. \ref{fig:housholdQoS} presents the QoS reduction for each customer in relation to the growth in HC, as depicted in Tab. \ref{tab:HCresults}. This reduction is achieved by employing the EV-NAHC approach across three separate daily charging energy scenarios. For instance, in the context of low-daily charging energy, the QoS is measured at a charging power of 9 kW, when the voltage incident is observed for EV-HC, and continues until 12 kW, when the aggregated QoS threshold is exceeded for EV-NAHC. The findings indicate that while the aggregated QoS threshold of the network is 0.8, the customers located farthest are experiencing a QoS of 0.6, which is significantly lower than the consumers located closer, who are not experiencing any alterations in their QoS. The impact of active network management on customers' QoS introduces a dimension of unfairness into the system's operation. Nevertheless, the EV-NAHC is augmented according to the aggregated QoS of all users.
This unfairness observed here will be reduced in our future work.

\vspace{15pt}

\begin{figure}[tp]
    \centering
    \includegraphics[width=0.95 \linewidth]{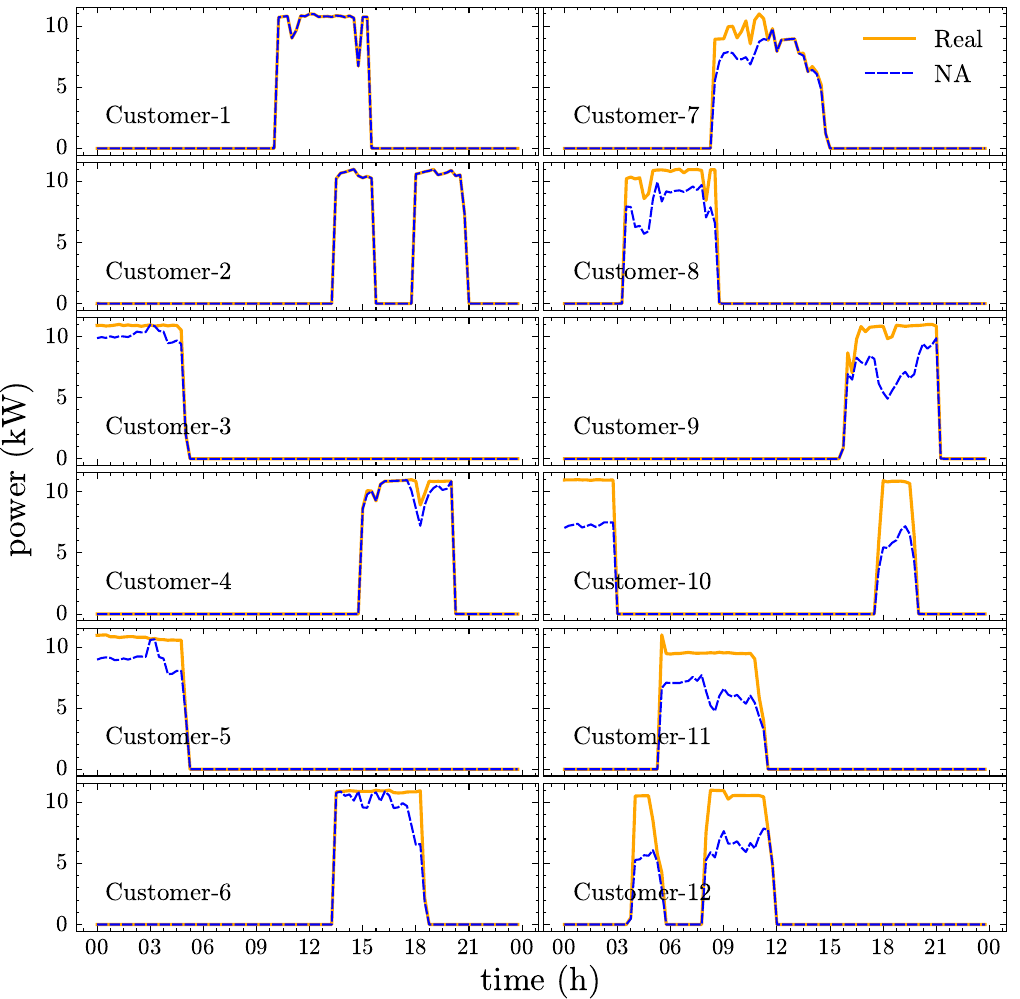}
    \vspace{-10pt}
    \caption{Charging power profiles of low-daily charging energy EVs: Real vs Network-aware}
    \label{fig:EVprofiles}
\end{figure}

\begin{figure}[!t]
    \centering
    \includegraphics[width=0.98 \linewidth]{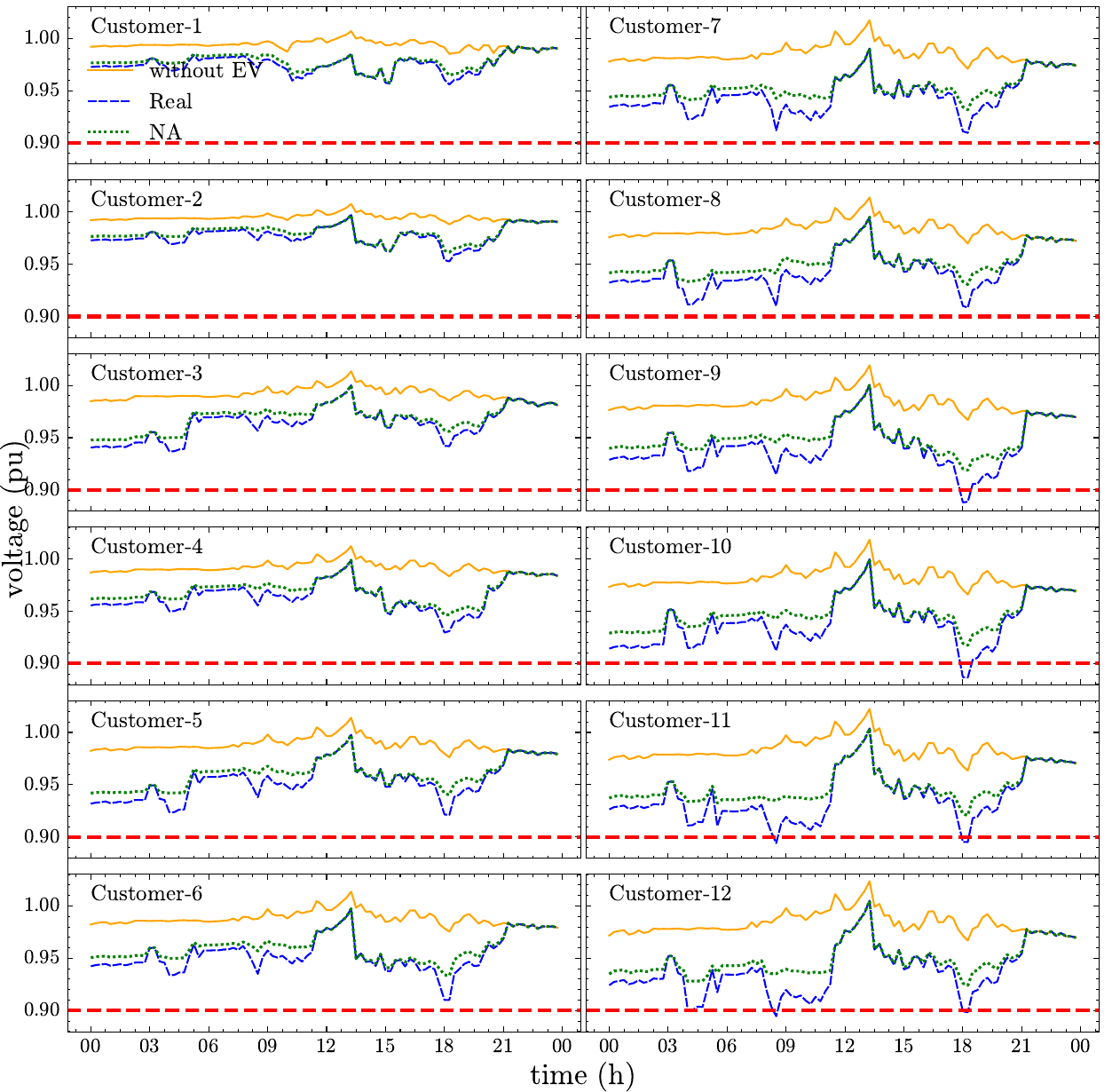}
    \vspace{-10pt}
    \caption{Customers voltage profiles for low-daily charging energy EVs scenario: Real vs Network-aware}
    \label{fig:voltageprofiles}
\end{figure}

\textbf{Evolution of charging trajectory:}
Fig. \ref{fig:EVprofiles} displays the original EV profiles and the updated profiles generated by DOE through active network management, which are used to validate the results of individual QoS and the EV-NAHC. Fig. \ref{fig:voltageprofiles} illustrates the voltage profiles of the consumers. Based on the findings, it can be inferred that users located further away from the transformer suffer a greater number of voltage issues. While active network management effectively resolves the issue of overvoltage for this user, their charging patterns will undergo significant modifications in comparison to other users. Consequently, their QoS will be significantly reduced compared to others.
Although the QoS reduction has a locational discrepancy, the EV-NAHC improved substantially for all cases.

\pagebreak

\section{Conclusion and future work} \label{conclusion}

This study developed an EV-NAHC framework that assesses the HC of an active distribution network by considering the aggregated QoS of all customers across three distinct daily charging scenarios for EVs. It has been demonstrated that DOEs can improve the EV-HC by maintaining the network operational parameters within predefined bounds. Nevertheless, an investigation into DOE generation highlighted the importance of defining their attributes ($\Delta_{\text{perm}}$, $factor$, aggregated QoS threshold) to mitigate the likelihood of unavoided DNIs recurring.

Conversely, it has been demonstrated that DOEs substantially decrease the QoS for some customers. Although the farthest users experience a greater deterioration in QoS, it was shown that the overall QoS might remain within the specified limit. Hence, to determine the HC of active distribution network for EVs, it was proposed to consider the aggregated QoS of all customers as a constraining factor. According to this new definition, the EV-HC is increased significantly for EVs with medium, and high-daily charging energy.

In future works, we aim to reduce the locational disparity of customers in the EV-NAHC framework.

\pagebreak

\section{Acknowledgements}

This work is supported by the Flemish Government and Flanders Innovation \& Entrepreneurship (VLAIO) through the Flux50 projects HUME (HBC.2021.0079) and IMPROcap (HBC.2022.0733).


\pagebreak

\bibliographystyle{IEEEtran}
\bibliography{reference.bib}

\end{document}